\documentstyle[epsfig]{elsart}
\newcommand{\lsim}{\mathrel{\rlap{\lower4pt\hbox{\hskip0pt$\sim$}}
\raise1pt\hbox{$<$}}}
\newcommand{\qcdtmu}{\mbox{QCD$_T^\mu$}}
\newcommand{\sfrac}[2]{\mbox{\footnotesize $\frac{#1}{#2}$}}
\begin{document}
\begin{frontmatter}
\hspace*{\fill}{Preprint Numbers: \parbox[t]{100mm}{ANL-PHY-8756-TH-97
        \hspace*{\fill} nucl-th/yymmddd \\ 
        MPG-VT-UR 106/97}}

\title{Thermodynamic properties of a simple, confining model}
\author[ur]{David Blaschke,}
\author[anl]{Craig D. Roberts}
\author[ur,israel]{and Sebastian Schmidt}
\address[ur]{Fachbereich Physik, Universit\"at Rostock, D-18051 Rostock,
Germany}
\address[anl]{Physics Division, Bldg. 203, Argonne National Laboratory,\\
Argonne IL 60439-4843, USA}
\address[israel]{School of Physics and Astronomy, Raymond and Beverly Sackler
Faculty of Exact Sciences, 
Tel Aviv University, 69978 Tel Aviv, Israel}
\begin{abstract}
We study the equilibrium thermodynamics of a simple, confining, DSE-model of
2-flavour QCD at finite temperature and chemical potential.  The model has
two phases: one characterised by confinement and dynamical chiral symmetry
breaking; and the other by their absence.  The phase boundary is defined by
the zero of the vacuum-pressure difference between the confined and
deconfined phases.  Chiral symmetry restoration and deconfinement are
coincident with the transition being of first order, except for $\mu=0$,
where it is second order.  Nonperturbative modifications of the dressed-quark
propagator persist into the deconfined domain and lead to a dispersion law
modified by a dynamically-generated, momentum-dependent mass-scale.  This
entails that the Stefan-Boltzmann limit for the bulk thermodynamic quantities
is attained only for large values of temperature and chemical potential.
\end{abstract}
\begin{keyword}
  Field theory at finite temperature and chemical potential; Confinement;
  Dynamical chiral symmetry breaking; Dyson-Schwinger equations\\[2mm]
\end{keyword}
\end{frontmatter}
%
The existence of a ``quark-gluon plasma'' [QGP] phase of QCD is well
established.  It is characterised by deconfinement, which means that quarks
and gluons have mean free paths that are long in comparison with standard
nuclear radii, and the realisation of chiral symmetry in the Wigner mode;
i.e., all chiral symmetry breaking effects require an explicit source: the
current-quark mass.  However, the details of the transition to this phase,
and the thermodynamic parameters that characterise it: energy, entropy,
pressure, etc.; remain uncertain.  These are the quantities that will
determine whether, and under what conditions, the QGP can be engineered in
contemporary experiments.

The thermodynamic properties of a theory are completely determined by its
partition function [generating functional], and this is the quantity
estimated directly in numerical simulations of lattice-QCD actions at
finite-$T$.  The Dyson-Schwinger equations~\cite{dserev} [DSEs] provide
another nonperturbative means of calculating the partition function: via the
evaluation of the Schwinger functions, which are the moments of the measure
that defines the partition function. Tractable, quantitative studies of the
coupled system of DSEs, examples of which are the QCD gap equation and meson
Bethe-Salpeter equation, require a truncation of the tower of integral
equations; for example, via a choice for the quark-antiquark scattering
kernel.  Although it is not possible to judge the fidelity of a particular
truncation {\it a priori}$\,$, a systematic approach~\cite{brs96} allows one
to address this issue in a direct, practical and constructive manner.  The
DSEs have been applied extensively at zero temperature and chemical potential
to the phenomenology of QCD~\cite{pct97}, including the study of strong
interaction contributions to weak interaction observables~\cite{weak}.

A robust, qualitative prediction of DSE studies in QCD is that, in the
infrared; i.e., for small spacelike-$q^2$, the propagation characteristics of
the elementary excitations are much altered from that inferred from
perturbation theory.  For example, the dressed-gluon propagator (2-point
function) is strongly enhanced~\cite{gpropir}, which leads via the QCD gap
equation to an infrared enhancement of the light-quark mass-function and, as
a result, an infrared suppression of the dressed-light-quark
propagator~\cite{dcsb}.  These modifications are intimately related to
confinement and dynamical chiral symmetry breaking and mean that new
challenges arise in applying the DSEs to QCD at finite-$T$ and $\mu$
[\qcdtmu].  Therefore, as a precursor to sophisticated DSE analyses of
\qcdtmu, it is useful to explore simple DSE-models in which these features
are manifest.

A particularly simple and useful illustrative DSE-model of QCD~\cite{MN83} is
specified by a model dressed-gluon propagator that can be generalised to
finite-$T$ as~\cite{BBKR96}
\begin{equation}
\label{mnprop}
g^2 D_{\mu\nu}(\vec{p},\Omega_k) = 
\left(\delta_{\mu\nu} 
- \frac{p_\mu p_\nu}{|\vec{p}|^2+ \Omega_k^2} \right)
2 \pi^3 \,\frac{\eta^2}{T}\, \delta_{k0}\, \delta^3(\vec{p})\,,
\end{equation}
where $(p_\mu)\equiv(\vec{p},\Omega_k)$, $\Omega_k = 2 k \pi T$ is the boson
Matsubara frequency\footnote{In our Euclidean formulation
$\{\gamma_\mu,\gamma_\nu\}=2\delta_{\mu\nu}$ with $\gamma_\mu^\dagger =
\gamma_\mu$.} and $\eta$ is a mass-scale parameter.  The infrared enhancement
of the dressed-gluon propagator suggested by Refs.~\cite{gpropir} is manifest
in this model.  As an infrared-dominant model that does not represent well
the behaviour of $D_{\mu\nu}(\vec{p},\Omega_k)$ away from $|\vec{p}|^2+
\Omega_k^2 \approx 0$, there are some model-dependent artefacts in our study.
However, there is significant merit in its simplicity and, since the
artefacts are easily identified, the model remains useful as a means of
elucidating easily many of the qualitative features of more sophisticated
Ans\"atze.

Using (\ref{mnprop}) and the ``rainbow-approximation'': $\Gamma_\mu^a(k,p)=
\sfrac{1}{2}\lambda^a \gamma_\mu$, for the dressed quark-gluon vertex; the
\qcdtmu\ gap equation, or DSE for the dressed-quark propagator,
is~\cite{brs96}
\begin{equation}
\label{mndse}
S^{-1}(\vec{p},\omega_k) = S_0^{-1}(\vec{p},\omega_k)
        + \sfrac{1}{4}\eta^2\gamma_\nu S(\vec{p},\omega_k) \gamma_\nu\,,
\end{equation}
where $S_0^{-1}(\vec{p},\omega_k)\equiv i\vec{\gamma}\cdot\vec{p}+i\gamma_4
(\omega_k+i\mu)+m$, $\omega_k = (2k + 1)\pi T$ is the fermion Matsubara
frequency, $\mu$ is the chemical potential and $m$ is the current-quark mass.
A simplicity inherent in (\ref{mnprop}) is now apparent: it allows the
reduction of an integral equation to an algebraic equation, in whose solution
many of the qualitative features of more sophisticated models are manifest,
as will become clear.  The solution of (\ref{mndse}) has the general form
\begin{equation}
\label{qprop}
S(\tilde p_k) = \frac{1}{i\vec{\gamma}\cdot\vec{p} A(\tilde p_k) 
+ i\gamma_4 (\omega_k+i\mu) C(\tilde p_k) 
+ B(\tilde p_k) }\,,
\end{equation}
where $\tilde p_k \equiv (\vec{p},\omega_k+i\mu)$, and (\ref{mndse}) entails
that the scalar functions introduced here satisfy
\begin{eqnarray}
\label{beqnfour}
\eta^2 m^2 & = & B^4 + m B^3 + \left(4 \tilde p_k^2 - \eta^2 -
        m^2\right) B^2 -m\,\left( 2\,{{\eta }^2} + {m^2} +
        4\,\tilde p_k^2 \right)B   \,,     \\ 
A(\tilde p_k) & = & C(\tilde p_k) = \frac{2 B(\tilde p_k)}{m +B(\tilde p_k)}\,.
\end{eqnarray}

Herein we are particularly interested in the chiral limit, $m=0$.  In this
case (\ref{beqnfour}) reduces to a quadratic equation for $B(\tilde p_k)$,
which has two qualitatively distinct solutions.  The ``Nambu-Goldstone''
solution, for which
\begin{eqnarray}
\label{ngsoln}
B(\tilde p_k) & = &\left\{
\begin{array}{lcl}
\sqrt{\eta^2 - 4 \tilde p_k^2}\,, & &\Re(\tilde p_k^2)<\sfrac{\eta^2}{4}\\
0\,, & & {\rm otherwise}
\end{array}\right.\\
C(\tilde p_k) & = &\left\{
\begin{array}{lcl}
2\,, & & \Re(\tilde p_k^2)<\sfrac{\eta^2}{4}\\
\sfrac{1}{2}\left( 1 + \sqrt{1 + \sfrac{2 \eta^2}{\tilde p_k^2}}\right)
\,,& & {\rm otherwise}\,,
\end{array}\right.
\end{eqnarray}
describes a phase or mode of this model in which: 1) chiral symmetry is
dynamically broken, because one has a nonzero quark mass-function, $B(\tilde
p_k)$, in the absence of a current-quark mass; and 2) the dressed-quarks are
confined, because the propagator described by these functions does not have a
Lehmann representation.  The alternative ``Wigner'' solution, for which
\begin{eqnarray}
\label{wsoln}
\hat B(\tilde p_k) & \equiv & 0\\
\hat C(\tilde p_k) & = &
\sfrac{1}{2}\left( 1 + \sqrt{1 + \sfrac{2 \eta^2}{\tilde p_k^2}}\right)\,,
\end{eqnarray}
describes a phase of the model in which chiral symmetry is not broken and the
dressed-quarks are not confined.  

With these two ``phases'', characterised by qualitatively different,
momentum-dependent modifications of the quark propagator, this DSE-model of
\qcdtmu\ can be used to explore chiral symmetry restoration and
deconfinement, and elucidate aspects of the methodology of such studies.

The pressure is obtained directly from the partition function, ${\cal Z}$,
which is the sum of all vacuum-to-vacuum transition amplitudes.  In
``stationary phase'' approximation, the partition function is given by the
tree-level auxiliary-field effective action~\cite{hay91} and the pressure is:
\begin{equation}
\label{lnz}
P[S] \equiv \frac{T}{V}\ln{\cal Z} = 
\frac{T}{V}\left\{ {\rm TrLn}\left[\beta S^{-1}\right]
        -\sfrac{1}{2} {\rm Tr}\left[\Sigma S\right]\right\}\,,
\end{equation}
where $\beta=1/T$ and the self energy $\Sigma(\tilde p_k) \equiv
S^{-1}(\tilde p_k) - S_0^{-1}(\tilde p_k)$.  The pressure is a functional of
$S(\tilde p_k)$.  In the absence of interactions $\Sigma \equiv 0$ and
(\ref{lnz}) yields the free fermion partition function.

We have neglected the gluon contribution to the pressure because, using
(\ref{mnprop}), it is a temperature independent constant.  More sophisticated
Ans\"atze do not suffer this defect~\cite{BBKR96,greg}, which is associated
with the fact that (\ref{mnprop}) does not represent well the ultraviolet
behaviour of $D_{\mu\nu}(k)$ in \qcdtmu.

The contribution to the partition function of hadrons and hadron-like
correlations is also neglected in (\ref{lnz}).  At the level of approximation
consistent with (\ref{lnz}) these terms are an additive contribution that can
be estimated using the {\it hadronisation} techniques of Ref.~\cite{cahill}.
After a proper normalisation of the partition function; i.e., subtraction of
the vacuum contribution, they are the only contributions to the partition
function in the domain of confinement.  They are easy to calculate and we
consider them no further herein.

\begin{figure}[t]
\centering{\
\epsfig{figure=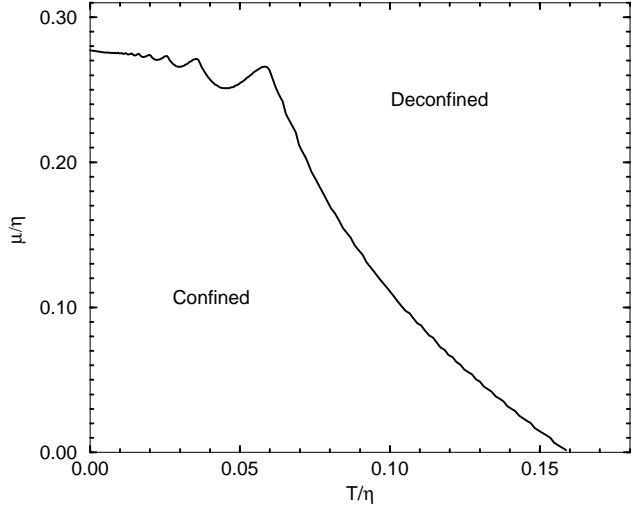,height=8.0cm}}
\caption{The phase boundary in the $(\bar T,\bar \mu)$-plane obtained from
(\protect\ref{bagconsgenA}) and (\protect\ref{bagcons});
$\eta=1.06\,$GeV~\protect\cite{MN83}.  The ``structure'' in this curve,
apparent for small-$T$, is an artefact of the fact that
(\protect\ref{mnprop}) improperly represents the quark-quark interaction in
the ultraviolet.
\label{critline}}
\end{figure}

In studying the phase transition one must consider the relative stability of
the confined and deconfined phases, which is measured by the
$(T,\mu)$-dependent vacuum pressure difference (or ``bag
constant''~\cite{cr85})
\begin{eqnarray}
\label{bagconsgen}
{\cal B}(T,\mu) & \equiv & P[S_{\rm NG}] - P[S_{\rm W}]\,, 
\end{eqnarray}
where $S_{\rm NG}$ means (\ref{qprop}) obtained from (\ref{ngsoln}) and
$S_{\rm W}$, (\ref{qprop}) obtained from (\ref{wsoln}).  ${\cal B}(T,\mu) >0$
indicates the stability of the confined (Nambu-Goldstone) phase and hence the
phase boundary is specified by that curve in the $(T,\mu)$-plane for which
\begin{eqnarray}
\label{bagconsgenA}
{\cal B}(T,\mu) & \equiv & 0\,.
\end{eqnarray}
In the chiral limit, the deconfinement and chiral symmetry restoration
transitions in this model are coincident.  The critical line is illustrated
in Fig.~\ref{critline}, where explicitly 
\begin{eqnarray}
\label{bagcons}
\lefteqn{{\cal B}(T,\mu) = }\\
&&\nonumber
 \eta^4\,2 N_c N_f \frac{\bar T}{\pi^2}\sum_{l=0}^{l_{\rm max}}
        \int_0^{\bar\Lambda_l}\,dy\,y^2\,
        \left\{\Re\left(2 \bar p_l^2 \right) 
                - \Re\left(\frac{1}{C(\bar p_l)}\right)
- \ln\left| \bar p_l^2 C(\bar p_l)^2\right|        \right\}\,,
\end{eqnarray}
with: $\bar T=T/\eta$, $\bar \mu=\mu/\eta$; and $\bar\omega^2_{l_{\rm
max}}\leq \sfrac{1}{4}+\bar\mu^2$, $\bar\Lambda^2 = \bar\omega^2_{l_{\rm
max}}-\bar\omega_l^2$, $\bar p_l = (\vec{y},\bar\omega_l+i\bar\mu)$.  

For $\mu=0$ the transition is second order and the critical temperature is
$T_c^0 = 0.159\,\eta$, which, using the value of $\eta=1.06\,$GeV obtained by
fitting the $\pi$ and $\rho$ masses~\protect\cite{MN83}, corresponds to
$T_c^0 = 0.170\,$GeV.  This is only 12\% larger than the value obtained in
Ref.~\cite{BBKR96} and the order of the transition is the same.  However, in
the present case the critical exponent is $\beta=0.5$, which differs from the
result $\beta \approx 0.33$ obtained in Ref.~\cite{BBKR96}.  This is an
artefact of the fact that (\ref{mnprop}) neglects short-range contributions
to the quark-quark interaction and, in particular, it is due to the sharp
``cut-off'' provided by $B(\tilde p_k)$ in (\ref{ngsoln}).  Nevertheless, the
comparison with Ref.~\cite{BBKR96} illustrates that this simple model can
provide a reasonable guide to the thermodynamic properties of more
sophisticated DSE-models of \qcdtmu.

\begin{figure}[t]
\centering{\
\epsfig{figure=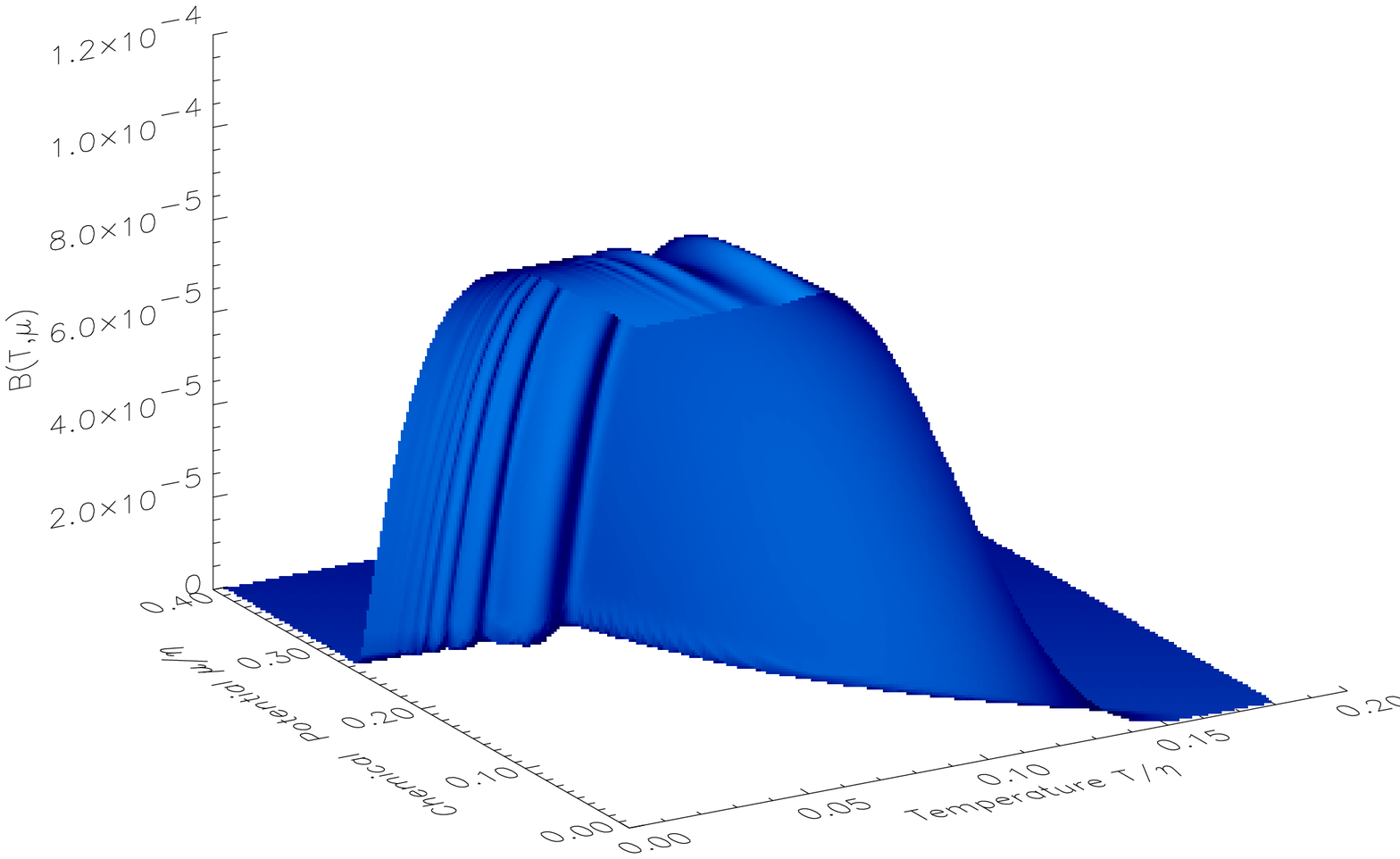,height=9.0cm}}
\caption{${\cal B}(T,\mu)$ from (\protect\ref{bagcons}); \mbox{${\cal
B}(T,\mu)>0$} marks the confinement domain.
\label{bagpres}}
\end{figure}

A feature of the simplicity of the model is that it allows a straightforward
analysis of $\mu\neq 0$, where numerical simulations of lattice-QCD actions
are not tractable~\cite{dks96}, thereby providing valuable insight into
deconfinement in this region.  For any $\mu \neq 0$ the transition is
first-order, as revealed by close scrutiny of Fig.~\ref{bagpres}.  For $T=0$
the critical chemical potential is $\mu_c^0=0.3\,$GeV.  We also see from
Fig.~\ref{critline} that $\mu_c(T)$ is insensitive to $T$ until $T\approx
\sfrac{1}{3}\,T_c^0$.

The $(T,\mu)$-dependent vacuum pressure difference, ${\cal B}(T,\mu)$, is
illustrated in Fig.~\ref{bagpres}.  The scale is set by \mbox{${\cal
B}(0,0)= (0.102\,\eta)^4 = (0.109\,{\rm GeV})^4$}, which can be compared with
the value $\sim (0.145\,{\rm GeV})^4$ commonly used in bag-like models of
hadrons~\cite{cahill}.
This pressure is associated with the rearrangement of the confined-quark
vacuum; i.e., with the rearrangement of the ``ground state''.  It does not
contribute actively to the thermodynamic pressure, describing only the change
in the ``stationary point'' that defines the auxiliary-field
effective-action.

Once in the deconfinement domain, illustrated clearly in
Fig.~\ref{critline}, the quarks contribute an amount 
\begin{equation}
\label{pwigner}
P[S_{\rm W}]= \eta^4\,2 N_c N_f \frac{\bar T}{\pi^2}\sum_{l=0}^{\infty}
        \int_0^{\infty}\,dy\,y^2\,
        \left\{\ln\left| \beta^2 \tilde p_l^2 \hat C(\bar p_l)^2\right|
        - 1 +  \Re\left(\frac{1}{\hat C(\bar p_l)}\right)  \right\}
\end{equation}
to the pressure, which we renormalise to zero on the phase boundary.  Just as
in the case of free fermions, this expression is formally divergent and one
must isolate and define the temperature-dependent, active contribution.  This
is made difficult by the fact that, in general, $\hat C(\bar p_l)$ is only
known numerically, and hence it is not possible to evaluate $P[S_{\rm W}]$
analytically.  We have developed a method for the numerical evaluation of
(\ref{pwigner}).

Consider the derivative of the integrand in (\ref{pwigner}):
\begin{eqnarray}
\label{presb}
\lefteqn{
\sum_{l=0}^\infty\,\frac{d}{d\bar T}\,\left\{\ln\left| \beta^2 \tilde p_l^2 \hat
        C(\bar p_l)^2\right| 
        - 1 +  \Re\left(\frac{1}{\hat C(\bar p_l)}\right)  \right\} = }\\
&& \nonumber
\sum_{l=0}^\infty\,
\left\{ -\frac{1}{\bar T} \left[
        \frac{(y-\bar\mu)^2}{(y-\bar\mu)^2+\bar\omega_l^2}
        + \frac{(y+\bar\mu)^2}{(y+\bar\mu)^2+\bar\omega_l^2}
        \right]
        +  \Re\left(
        \frac{ 2 \hat C(\bar p_l) - 1}{\hat C(\bar p_l)^2}\,
        \frac{d \hat C(\bar p_l)}{d \bar T\;\;\;\;}\right)
\right\}\,.
\end{eqnarray}
In the absence of interactions $C(\bar p_l)\equiv 1$, the second term is zero
and
\begin{equation}
\label{presa}
-\frac{2}{\bar T} \sum_{l=0}^\infty\,
\left[\frac{(y-\bar\mu)^2}{(y-\bar\mu)^2+\bar\omega_l^2}
        + \frac{(y+\bar\mu)^2}{(y+\bar\mu)^2+\bar\omega_l^2}\right]  
= \frac{d}{d\bar T}\left\{
        \frac{e(y)}{\bar T} + {\cal I}(e(y))\right\}\,,
\end{equation}
where in this case $e(y)=y$ and 
\begin{equation}
{\cal I}(\zeta)  = 
        \ln\left[1 + \exp\left(-\frac{\zeta-\bar\mu}{\bar T}\right)\right]
        + \ln\left[1 + \exp\left(-\frac{\zeta+\bar\mu}{\bar T}\right)\right]\,.
\end{equation}
Appropriately inserting (\ref{presa}) for the parenthesised term in
(\ref{pwigner}), and neglecting $T$-independent terms one obtains,
\begin{eqnarray}
P[S_0] & = & \eta^4\,N_c N_f \frac{\bar
        T}{\pi^2}\int_0^{\infty}\,dy\,y^2\, {\cal I}(y)\\
 & = & \label{sbpres}
        \eta^4\, N_c N_f \frac{1}{12\pi^2}\left(
        \bar\mu^4 + 2 \pi^2 \bar\mu^2 \bar T^2 + \frac{7}{15}\pi^4
                        \bar T^4\right)\,, 
\end{eqnarray}
which is the massless, free particle pressure.

To proceed we assume that the nontrivial momentum dependence of $\hat C(\bar
p_l)$, which is manifest in all DSE-models of QCD, acts primarily to modify
the usual massless, free particle dispersion law.  We evaluate numerically
the sum on the right-hand-side of (\ref{presb}) and use the form on the
right-hand-side of (\ref{presa}) to fit a modified, $T$-independent
dispersion law, $\underline{e}(y,\bar\mu)= y +\kappa(y,\bar\mu)$, to the
numerical results.  The existence of a $\kappa(y,\bar\mu)$ that provides a
good $\chi^2$-fit on the deconfinement domain is understood as an {\it a
posteriori} justification of this assumption.  In our calculations, on the
entire $T$-domain, the relative error between the fit and the numerical
results is $< 10$\%.  

\begin{figure}[t]
\centering{\
\epsfig{figure=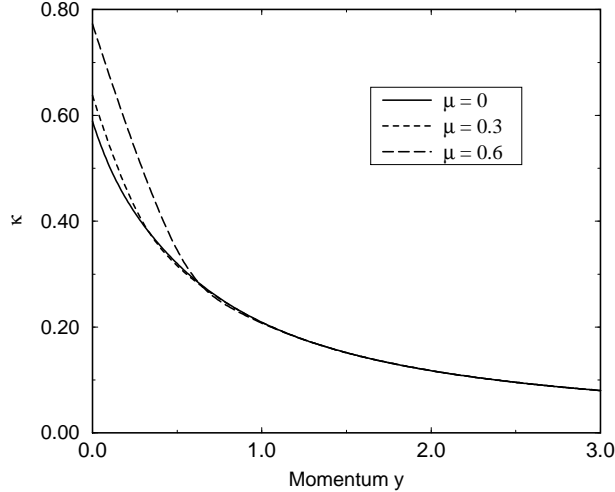,height=9.0cm,angle=-90}}
\caption{$\kappa(y,\bar\mu)$, which describes the nonperturbative
modification of the free particle dispersion law, for $\bar\mu=0, 0.3, 0.6$.
By assumption, it is independent of $T$.
\label{figrho}}
\end{figure}
We illustrate the calculated form of $\kappa(y,\bar\mu)$ in
Fig.~\ref{figrho}; it only depends weakly on $\bar\mu$.  The form indicates
that the result of the persistence of nonperturbative effects into the domain
of deconfinement; i.e., the nontrivial momentum dependence of $\hat C(\bar
p_l)$ and its slow evolution to the asymptotic value $\hat C(\bar p_l)=1$, is
to generate a mass-scale in the massless dispersion law: $\kappa(0,0)\simeq
0.6 \sim 2 \bar\mu_c^0$.  This mass-scale is unrelated to the chiral-symmetry
order parameter, $B(\vec{0},\omega_0+i\mu)$, and is a qualitatively new
feature of this study.  For $\bar\mu> 5 \bar\mu_c^0$ the explicit mass-scale
introduced by the chemical potential overwhelms this dynamically generated
scale.

With this result (\ref{pwigner}) becomes
\begin{eqnarray}
P[S_{\rm W}] & = & 
\eta^4\,N_c N_f \frac{\bar
        T}{\pi^2}\int_0^{\infty}\,dy\,y^2\, 
        {\cal I}(\underline{e}(y,\bar\mu))\,,
\end{eqnarray}
and the quark pressure in this DSE-model of \qcdtmu\ is 
\begin{equation}
\label{qpres}
P_q(T,\mu)  =  \theta({\cal D})\left\{
        P[S_{\rm W}] - \left.P[S_{\rm W}]\right|_{\partial {\cal D}}\right\}
\end{equation}
where ${\cal D}$ is the domain marked ``Deconfined'' in Fig.~\ref{critline},
$\theta( {\cal D})$ is a step function, equal to one for $(T,\mu)\in {\cal
D}$, and $\left.P[S_{\rm W}]\right|_{\partial {\cal D}}$ indicates the
evaluation of this expression on the boundary of ${\cal D}$, as defined by
the intersection of a straight-line from the origin in the $(T,\mu)$-plane to
the argument-value.
\begin{figure}[t]
\centering{\
\epsfig{figure=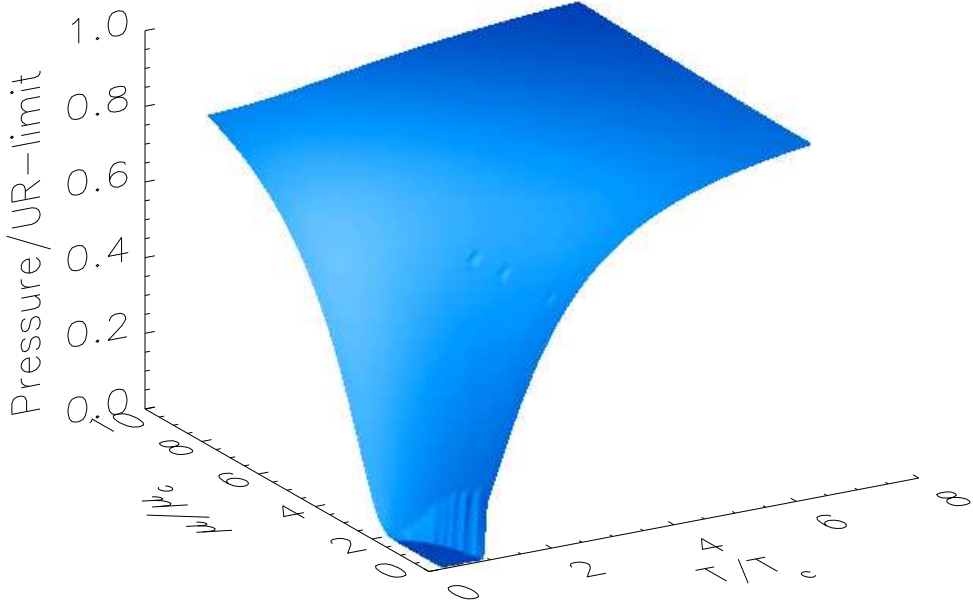,height=9.0cm}}
\caption{The quark pressure, $P_q(\bar T,\bar\mu)$, normalised to the free,
massless (or Ultra-Relativistic) result, (\protect\ref{sbpres}).
\label{presfig}}
\end{figure}
It is plotted in Fig.~\ref{presfig}, which illustrates clearly that in this
model the free particle (Stefan-Boltzmann) limit is reached at large values
of $\bar T$ and $\bar\mu$.  The approach to this limit is slow, however.  For
example, at $\bar T \sim 0.3 \sim 2 \bar T_c^0$, or $\bar \mu \sim 1.0 \sim 3
\bar\mu_c^0$, (\ref{qpres}) is only $\sfrac{1}{2}$ of the free particle
pressure, (\ref{sbpres}).  A qualitatively similar result is observed in
numerical simulations of lattice-QCD actions at finite-$T$~\cite{karsch97}.
This feature results from the slow approach to zero with $y$ of
$\kappa(y,\bar\mu)$, illustrated in Fig.~\ref{figrho}, and emphasises the
persistence of the momentum dependent modifications of the quark propagator.

\begin{figure}[t]
\centering{\
\epsfig{figure=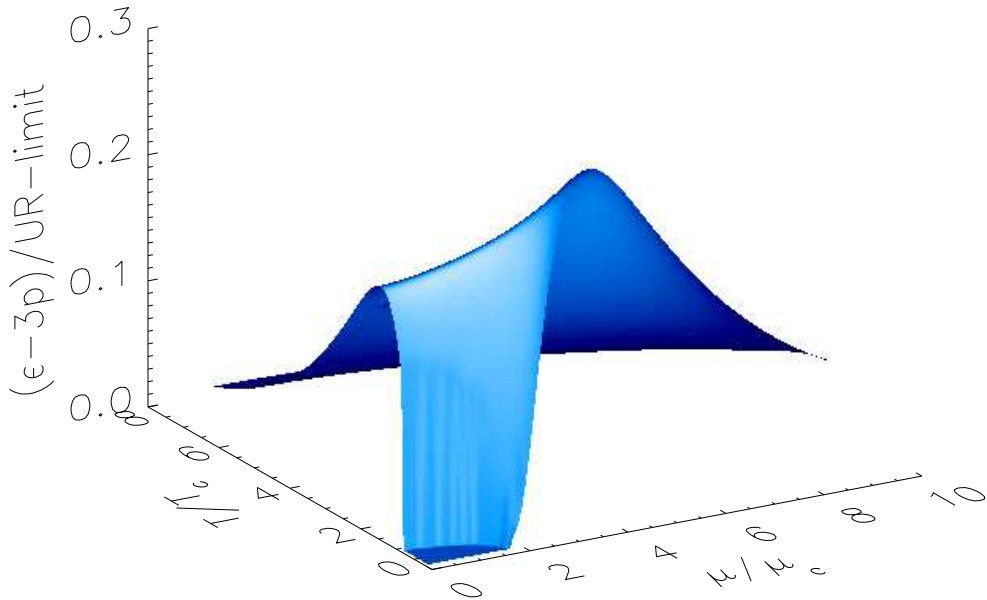,height=9.0cm}}
\caption{The ``interaction measure'', $\Delta(T,\mu)$, normalised to the
free, massless result for the pressure, (\protect\ref{sbpres}).
\label{energy}}
\end{figure}

With the definition and calculation of the pressure, $P_q(T,\mu)$, or
equivalently the partition function, all of the remaining bulk thermodynamic
quantities that characterise our DSE-model of \qcdtmu\ can be calculated.  As
an example, in Fig.~\ref{energy} we plot the ``interaction measure'':
$\Delta\equiv \epsilon - 3 P$, where $\epsilon$ is the energy density.  This
is zero for an ideal gas, hence the nomenclature: $\Delta$ measures the
interaction-induced deviation from ideal gas behaviour.
This figure provides a very clear indication of the persistence of
nonperturbative effects into the deconfinement domain.

We have explored the equilibrium thermodynamics of a simple, confining,
DSE-model of \qcdtmu, defined by the model dressed-gluon propagator in
(\ref{mnprop}), which shares many of the features found in DSE-models with
more realistic Ans\"atze for the dressed-gluon propagator~\cite{BBKR96}.  The
simplicity of the model allows a straightforward analysis of the bulk
thermodynamics for finite chemical potential as well as finite temperature.

Deconfinement and chiral symmetry restoration are coincident and the phase
boundary, defined by that curve in the $(T,\mu)$-plane for which the
temperature- and chemical-potential-dependent ``bag constant'',
(\ref{bagcons}), vanishes, is illustrated in Fig.~\ref{critline}.  The
transition is first order except for $\mu=0$.  We expect these results to be
qualitatively reliable; a more sophisticated model leads only to a slight
reduction in the $\mu=0$ critical temperature~\cite{BBKR96} and an increase
in the $T=0$ critical chemical potential~\cite{greg}.

The quark pressure is given by (\ref{qpres}) and is illustrated in
Fig.~\ref{presfig}.  It is zero in the confining domain and approaches the
Stefan-Boltzmann limit for a massless, free particle only for relatively
large values of $T$ and $\mu$.  The calculation of the pressure is
complicated by the momentum dependence of the nonperturbatively-dressed quark
propagator.  This generates a nonperturbative mass-scale in the dispersion
relation for the deconfined quark, Fig.~\ref{figrho}.  It is an essential
feature of this study and is responsible for the slow attainment of the
Stefan-Boltzmann limit.

The persistence of nonperturbative effects into the deconfining domain is
elucidated in the ``interaction measure'', which is illustrated as a function
of $T$ and $\mu$ in Fig.~\ref{energy}.  The slow attainment of the
Stefan-Boltzmann limit at finite-$T$ observed in lattice simulations of
\qcdtmu\ is mirrored in a slow approach to this limit with increasing $\mu$.

We acknowledge useful conversations with A. H\"oll and P. Maris.  DB and SS
gratefully acknowledge the hospitality of the Physics Division at ANL; and
CDR that of the Max-Planck-Group and the Department of Physics at the
University of Rostock.  CDR acknowledges a stipend from the
Max-Planck-Gesellschaft; and SS, financial support from the MINERVA
Foundation.  This work was supported in part by the Deutscher Akademischer
Austauschdienst; by the Department of Energy, Nuclear Physics Division, under
contract no. W-31-109-ENG-38; by the National Science Foundation under grant
no. INT-9603385; and benefited from the resources of the National Energy
Research Scientific Computing Center.



\begin{thebibliography}{99}
%
\bibitem{dserev} C. D. Roberts and A. G. Williams,
Prog. Part. Nucl. Phys. {\bf 33} (1994) 477.
%
\bibitem{brs96} A. Bender, C. D. Roberts and L. v. Smekal,
Phys. Lett. B {\bf 380} (1996) 7.
%
\bibitem{pct97} P.~C.~Tandy, ``Hadron physics from the Global Colour Model of
QCD'', archive: nucl-th/9705018, to appear in Prog. Part. Nucl. Phys. {\bf
39} (1997); M.~A.~Pichowsky and T.-S. H. Lee, ``Exclusive diffractive
processes and the quark substructure of mesons'', archive: nucl-th/9612049,
to appear in Phys. Rev. D.
%
\bibitem{weak} Yu. L. Kalinovsky, K. L. Mitchell and C. D. Roberts,
Phys. Lett. B {\bf 399} (1997) 22; M. B. Hecht and B. H. J. McKellar,
``Dipole Moments of the Rho Meson'', archive: hep-ph/9704326.
%
\bibitem{gpropir} M. Baker, J. S. Ball and F. Zachariasen, Nucl. Phys. B {\bf
186} (1981) 531; {\it ibid}, 560; D. Atkinson, {\it et al}., Nuovo Cimento A
{\bf 77} (1983) 197; N. Brown and M. R. Pennington, Phys. Rev. D {\bf 39}
(1989) 2723.
%
\bibitem{dcsb} A. G. Williams, G. Krein and C. D. Roberts, Ann. Phys. {\bf
210} (1991) 464; H.~J.~Munczek and P. Jain, Phys. Rev. D {\bf 46} (1992) 438.
%
\bibitem{MN83} H. J. Munczek and A. M. Nemirovsky, Phys. Rev. D {\bf 28}
(1983) 181.  
%
\bibitem{BBKR96} A. Bender, {\it et al}., Phys. Rev. Lett. {\bf 77} (1996)
3724.
%
\bibitem{hay91} R. W. Haymaker, Riv. Nuovo Cim. {\bf 14}, series 3, no. 8
(1991). 
%
\bibitem{greg} G. Poulis, C. D. Roberts and A. W. Thomas, in progress.
%
\bibitem{cahill} R. T. Cahill, Aust. J. Phys. {\bf 42} (1989) 171.
%
\bibitem{cr85} R. T. Cahill and C. D. Roberts, Phys. Rev. D {\bf 32} (1885) 2419.
%
\bibitem{dks96} M.-P. Lombardo, J. B. Kogut and D. K. Sinclair, Phys. Rev. D
{\bf 54} (1996) 2303.
%
\bibitem{karsch97} J. Engels, {\it et al}., Phys. Lett. B {\bf 396} (1997)
210.
%
\end{thebibliography}
\end{document}